\documentclass[10pt,reprint,amsmath,amssymb,floatfix,showpacs]{revtex4-1}
\usepackage{subfig} 
\usepackage{graphicx}
\usepackage{epsfig}
\usepackage{bm}
\usepackage{float}
\usepackage{subfloat}
\usepackage{dcolumn}
\begin{document}
\preprint{APS/123-QED}
\title{Charged Higgs boson contribution to $\overline{\nu}_e-e$ scattering from low to ultrahigh energy in Higgs Triplet model}
\author{Jian Fuh Ong, Ithnin Abdul Jalil and Wan Ahmad Tajuddin Wan Abdullah}
\affiliation{Department of Physics, Faculty of Science, University of Malaya, 50603 Kuala Lumpur,
Malaysia.}
\date{\today}
\begin{abstract}
We study the $\overline{\nu}_e-e$ scattering from low to
ultrahigh energy in the framework of Higgs Triplet Model (HTM). We add the contribution of charged Higgs boson exchange to the total cross section of the scattering. We obtain the upper bound $h_{ee}/M_{H^\pm}\lesssim2.8\times10^{-3}GeV^{-1}$ in this process from low energy experiment. We show that by using the upper bound obtained, the charged Higgs contribution can give enhancements to the total cross section with respect to the SM prediction up to 5.16 $\%$ at $E\leq10^{14}$ eV and maximum at $s\approx M_{H^\pm}^2$ and would help to determine the feasibility experiments to discriminate between SM and HTM at current available facilities.
\end{abstract}
\maketitle
\section{Introduction}
In the standard model (SM), charged leptons, quarks and gauge bosons  obtained their masses through Higgs mechanism \cite{PhysRevLett.13.321, *PhysRevLett.13.508, *PhysRevLett.13.585} after spontaneous symmetry breaking \cite{RevModPhys.81.1015} while neutrinos remain massless. However, studies on solar neutrino problem \cite{Bahcall23011976} and neutrino oscillation \cite{Bilenky1978225} have led physicists to suspect that the neutrino is massive. The discovery of massive neutrino have motivated the study of neutrino to the physics beyond SM. In particular, one of the simplest extensions is the Higgs triplet model (HTM) \cite{PhysRevD.22.2860, *PhysRevD.22.2227}. This model provides us an alternative way to introduce and explain the smallness of the neutrinos masses
through type-II seesaw mechanism and
enables one to study the lepton violation processes \cite{PhysRevD.78.015018}. The introduction of singly and doubly charged Higgs bosons in this model have opened many phenomenological studies in hunting for them at high energy frontier accelerators particularly at the CERN LHC and Tevatron \cite{PhysRevD.78.015018,
PhysRevD.72.035011, *PhysRevD.84.035028, *PhysRevD.84.035010}. The direct search for $H^{\pm\pm}$ had been carried out at Tevatron by assuming the process $p\overline{p}\rightarrow H^{++}H^{--}\rightarrow l^+l'^+l^-l'^-$, and the mass limit $M_{H^{\pm\pm}}>$ 118 GeV were derived \cite{PhysRevLett.93.141801, *PhysRevLett.93.221802}.

Neutrino-electron elastic scattering is a pure leptonic process that
provide precise test to the Standard model (SM) of electroweak theory. This reaction proceeds through charge current (CC), neutral current (NC) and their interference. 
The neutrino-electron elastic scattering has been widely studied in experiment by using $\nu_e (%
\overline{\nu}_e)$ beam up to several MeV \cite{PhysRevD.63.112001,
PhysRevD.81.072001} and consistent with the SM prediction. On the other hand, the study of these processes to higher energy 
($10^{12}$ eV $\leq E \leq$ $10^{20}$ eV) have been extensively discussed by Glashow, Mikaelian and Zheleznykh \cite
{PhysRev.118.316, PhysRevD.22.2122} in the framework of SM. At this range of energy the resonance production of $W^-$ boson or Glashow resonance 
\cite{PhysRev.118.316} is expected via $\overline{\nu}_e-e$ annihilation. However, the experiment evidences for Glashow resonance are not yet confirmed at present and the physics may be tested by using large scale neutrino detectors in the next few years.

Neutrino telescope such as IceCube \cite{10.11402} is used to detect the ultrahigh energy (UHE) cosmic neutrinos from extraterrestrial
sources \cite{V1995363}. The flux of ultrahigh energy $\overline{\nu}_e$ may get enhanced through the neutrino oscillation of $\overline{\nu}_\mu$. The study of ultrahigh energy particles interaction would be crucial to probe the feasibility of physics beyond the standard model. 

In this paper, we study the $\overline{\nu}_e-e$ scattering by adding the contribution of charged Higgs boson exchange. Our interest is to study how much the charged Higgs boson exchange contributes to the total cross section from low to ultrahigh energy in HTM. In HTM, the coupling of charged leptons to neutrinos are proportional to $h_{ll'}$. Therefore, for small $\upsilon_\Delta$ and large $h_{ll'}$ while keeping the neutrino mass below the experimental upper bound the contribution of charged Higgs boson exchange can be significant. On the contrary, the couplings of charged Higgs bosons to charged leptons and neutrinos are proportional to $m_f \times \tan\beta$ in two Higgs doublet model of type II (2HDM(II)) \cite{gunion2000higgs}. The lower mass limits of charged Higgs boson in this model are  $M_{H^\pm}> 79.3$ GeV and $M_{H^\pm}\gtrsim 1.71 \tan\beta$ GeV which imply that $\tan\beta\gtrsim46$ where it is out of our interest at low energy regime \cite{springerlink:10.1140/epjc/s2005-02370-2,Heister20021}.

The outline of the paper is as follows: We present the the HTM briefly in  Sec. II. The total cross section of $\overline{\nu}_e-e$ scattering with the contribution from charged Higgs boson exchange is presented in Sec. III. In Sec. IV, the upper bound $h_{ee}/M_{H^\pm}$ is estimated. Result and discussion are discussed in Sec. V. Conclusions are given in Sec. VI.

\section{Higgs Triplet Model}
In Higgs triplet model, a $\mathit{I}$=1, $\mathit{Y}$=2 complex $%
SU(2)_L $ scalar triplet is included to the SM Lagrangian to explain the
smallness of neutrino mass \cite{PhysRevD.22.2860, *PhysRevD.22.2227} without
requiring extra right-handed neutrinos. The $SU(2)_L \times U(1)_Y$ gauge invariance Yukawa Lagrangian
for the interaction of leptons with scalar triplet is written as \cite{PhysRevD.78.015018, PhysRevD.81.115007} 
\begin{equation}
\mathcal{L}=h_{ll'}\psi^T_{l L}Ci\sigma_2\Delta\psi_{l' L} + H.c \label{eq1}
\end{equation}
where $\mathit{{h_{ll'}}(l,l'=e,\mu,\tau)}$ is the coupling constant matrix, C
is the charged conjugation matrix, $\sigma_2$ is a Pauli matrix, $%
\psi_{l L}=(\nu_l,l)^T_L$ is left-handed lepton doublet and $\Delta$ is 2$%
\times$2 representation of the $\mathit{Y}$ = 2 complex triplet field, 
\begin{equation}
\Delta= \left( 
\begin{array}{cc}
\Delta^+/\sqrt{2} & \Delta^{++} \\ 
\Delta^0 & - \Delta^+/\sqrt{2}
\end{array}
\right) \label{eq2}
\end{equation}
The Higgs triplet potential is
\begin{eqnarray}
V(\phi ,\Delta )&=&-m_H^2\phi ^\dagger \phi+M_{\Delta }^2Tr\Delta ^\dagger \Delta +\frac{\lambda}{4}(\phi ^\dagger \phi)^2\nonumber\\&&+\left ( \mu \phi ^Ti\sigma _2\Delta ^\dagger \phi +H.c. \right )\nonumber\\
&&+\lambda _{1}(\phi ^\dagger \phi )Tr\Delta ^\dagger \Delta +\lambda _2(Tr\Delta ^\dagger \Delta )^2\nonumber\\&&+\lambda _3Tr(\Delta ^\dagger \Delta )^2+\lambda _4\phi ^\dagger \Delta \Delta ^\dagger \phi \label{eq2.35}
\end{eqnarray}
where $\phi=(\phi^0,\phi^\pm)^T$ is the SM Higgs doublet.
The neutrinos acquire their mass after $\Delta^0$  gets the vacuum
expectation value (vev), $\upsilon_\Delta$ 
\begin{equation}
m_{\nu}=2h_{ll}\langle\Delta^0\rangle=\sqrt{2}h_{ll}\upsilon_\Delta
\label{eq3}
\end{equation}
On the other hand, the neutral components of Higgs doublet and triplet are expressed as
\begin{eqnarray}
\phi ^{0}&=&\frac{\upsilon _{0}+h^0+i\xi^0 (x)}{\sqrt{2}}\nonumber\\
\Delta^{0}&=&\frac{\upsilon _{\Delta}+\delta^0+i\eta^0 (x)}{\sqrt{2}}\label{eq2.37}
\end{eqnarray}
The ground state of a system where it is at the most stable state can be obtained by minimization the potential
\begin{eqnarray}
\frac{\partial V}{\partial \phi }_{\phi =\left \langle \phi \right \rangle, \Delta =\left \langle \Delta \right \rangle}&=&0\nonumber\\
\frac{\partial V}{\partial \Delta}_{\phi =\left \langle \phi \right \rangle, \Delta =\left \langle \Delta \right \rangle}&=&0\label{eq2.38}
\end{eqnarray}
we obtain
\begin{eqnarray}
&&-m_H^2+\lambda \frac{\upsilon ^2_0}{4}-\sqrt{2}\mu\upsilon _\Delta +(\lambda _1+\lambda_4)\frac{\upsilon _\Delta ^2}{2}=0
\end{eqnarray}
and
\begin{eqnarray}
&&M^2_\Delta \upsilon _\Delta -\frac{\mu\upsilon ^2_0}{\sqrt{2}}+\frac{1}{2}(\lambda _1+\lambda _4)\upsilon ^2_0\upsilon _\Delta ^2+(\lambda _2+\lambda _3)\upsilon _\Delta ^3=0\nonumber\\\label{eq2.39}
\end{eqnarray}
By considering the lowest order processes, the $\lambda_i$ parameters are set to zero. Thus we have
\begin{eqnarray}
\upsilon_{\Delta}=\frac{\mu\upsilon_0^2}{\sqrt{2}M_{\Delta}^2}, ~m_H^2=\left(\frac{\lambda }{4}-\frac{\mu ^{2}}{M_\Delta ^2}\right)\upsilon _0^2
\end{eqnarray}
From the Higgs potential one finds the masses of two CP-even, one CP-odd  and two singly charged Higgs bosons which are mixed of weak isospin doublet and triplet. Meanwhile, the doubly charged Higgs bosons are composed of triplet alone. The mass-squared matrix for CP-even states are
\begin{eqnarray}
\frac{1}{2}\left (\begin{array}{c c} h^0 & \delta^0\end{array}\right )\left(\begin{array}{c c}
\frac{\lambda\upsilon _0^2}{2} & -\sqrt{2}\mu \upsilon _0\\ -\sqrt{2}\mu \upsilon _0
& M_\Delta ^2
\end{array}\right)\left(\begin{array}{c c}
h^0\\ \delta^0\end{array}\right)\label{eq2.40}
\end{eqnarray}
Upon diagonalization, the two physical CP-even states are
\begin{eqnarray}
H_1&=&\cos \theta _0~h^0+\sin \theta _0~\delta ^0\nonumber\\
H_2&=&-\sin\theta_0~h^0+\cos\theta_0~\delta^0\label{eq2.41}
\end{eqnarray}
with the masses
\begin{equation}
M_{H_1}^2\approx \frac{\lambda \upsilon _0^2}{2 }-2\sqrt{2}\mu \upsilon _\Delta
,~~M_{H_2}^2\approx M_\Delta ^2+2\sqrt{2}\mu \upsilon _\Delta\label{eq2.42}
\end{equation}
and mixing angle
\begin{equation}
\tan 2\theta_0=-\frac{4M_\Delta ^2\upsilon _\Delta }{\upsilon _0(M_{H_1}^2+M_{H_2}^2-2M_\Delta ^2)}\label{eq2.43}
\end{equation}
The mass-squared matrix for CP-odd states are
\begin{eqnarray}
&&\frac{1}{2}\left ( \begin{array}{c c} \xi ^0 & \eta^0\end{array} \right )\left (\begin{array}{c c}2\sqrt{2}\mu \upsilon _\Delta & -\sqrt{2}\mu \upsilon _0 \\ -\sqrt{2}\mu \upsilon _0 & M_\Delta ^2\end{array} \right )\left(\begin{array}{c c} \xi ^0 \\ \eta^0\end{array}\right )\nonumber\\&&=
\frac{1}{2}\left ( \begin{array}{c c} G^0 & A^0\end{array} \right )\left (\begin{array}{c c}0 & 0 \\ 0 & M_\Delta ^2+2\sqrt{2}\mu \upsilon _\Delta \end{array} \right )\left(\begin{array}{c c} G ^0 \\ A ^0\end{array}\right )\label{eq2.44}
\end{eqnarray}
with 
\begin{eqnarray}
G^0&=&\cos \alpha~ \xi ^0+\sin \alpha ~\eta^0\nonumber\\
A^0&=&-\sin\alpha~\xi^0+\cos\alpha~\eta^0\label{eq2.45}
\end{eqnarray}
and
\begin{eqnarray}
M_{A^0}^2=M_\Delta ^2+2\sqrt{2}\mu \upsilon _\Delta,~M^2_{G^0}=0 
\end{eqnarray}
\begin{eqnarray}
\cos \alpha =\frac{\upsilon _0}{\sqrt{\upsilon _0^2+4\upsilon _\Delta ^2}},~\sin \alpha =\frac{2\upsilon _\Delta }{\sqrt{\upsilon _0^2+4\upsilon _\Delta ^2}}\label{eq2.46}
\end{eqnarray}
Meanwhile, the mass-squared matrix of the singly charged Higgs bosons are
\begin{eqnarray}
&&\left(\begin{array}{c c}\phi^+ & \delta^+ \end{array} \right )\left ( \begin{array}{c c} \sqrt{2}\mu\upsilon_\Delta & -\mu\upsilon_0 \\-\mu\upsilon_0 & M_\Delta^2\end{array} \right )\left ( \begin{array}{c c}\phi^+ \\ \delta^+ \end{array} \right )\nonumber\\&&=
\left(\begin{array}{c c}G^+ & H^+ \end{array} \right )\left ( \begin{array}{c c} 0 & 0 \\ 0 & M_\Delta^2+\sqrt{2}\mu\upsilon_\Delta\end{array} \right )\left ( \begin{array}{c c}G^+ \\ H^+ \end{array} \right )\label{eq2.47}
\end{eqnarray}
with
\begin{eqnarray}
G^\pm&=&\cos \theta_\pm~ \phi^\pm+\sin \theta_\pm ~\delta^\pm\nonumber\\
H^\pm&=&-\sin\theta_\pm~ \phi^\pm+\cos\theta_\pm~ \delta^\pm\label{eq2.48}
\end{eqnarray}
and
\begin{eqnarray}
M_{H^\pm}^2=M_\Delta ^2+\sqrt{2}\mu \upsilon _\Delta,~M^2_{G^\pm}=0 
\end{eqnarray}
\begin{eqnarray}
\cos \theta _+ =\frac{\upsilon _0}{\sqrt{\upsilon _0^2+2\upsilon _\Delta ^2}},~~~\sin \theta _+ =\frac{\sqrt{2}\upsilon _\Delta }{\sqrt{\upsilon _0^2+2\upsilon _\Delta ^2}}\label{eq2.49}
\end{eqnarray}
The doubly charged Higgs bosons are 
\begin{equation}
H^{\pm\pm}=\Delta^{\pm\pm},~M^2_{H^{\pm\pm}}=M_\Delta^2\label{eq2.50}
\end{equation}
The Goldstone bosons $G^0$ and $G^\pm$ are being absorbed by $Z^0$ and $W^\pm$ to acquire mass. The upper bound of $\upsilon_\Delta/\upsilon_0$ is set by $\rho=1.0008^{+0.0017}_{-0.0007}$ \cite{09543899377A075021} to be $\upsilon_\Delta/\upsilon_0 \lesssim0.02$. This imply that the mixing between Higgs doublet and triplet is negligible and therefore
\begin{equation}
M_{H^{\pm }}\approx M_{H_2}\approx M_{A^0}\approx M_{H^{\pm \pm }}=M_{\Delta}\label{eq3.64}
\end{equation}

\section{Total Cross Section}
\begin{figure}[H]
\subfloat[]{\includegraphics[width=4.0cm,height=3.0cm]{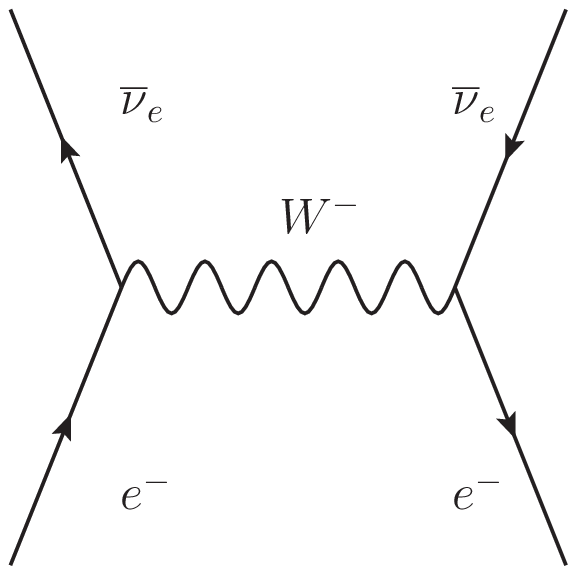}\label{figF1}}
\subfloat[]{\includegraphics[width=4.0cm,height=3.0cm]{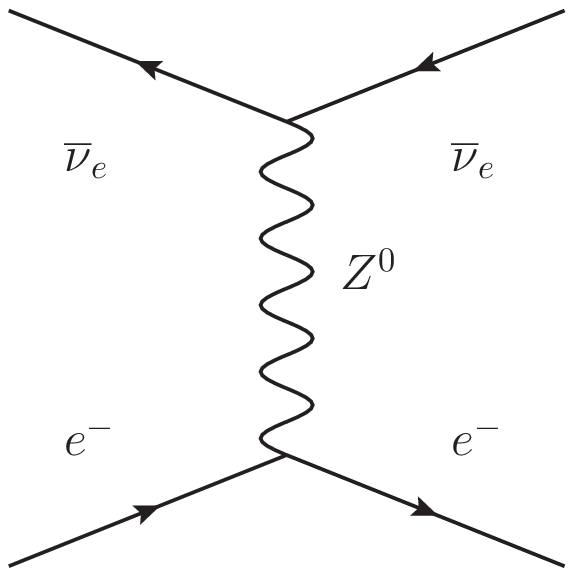}\label{figF2}}\\
\subfloat[]{\includegraphics[width=4.0cm,height=3.0cm]{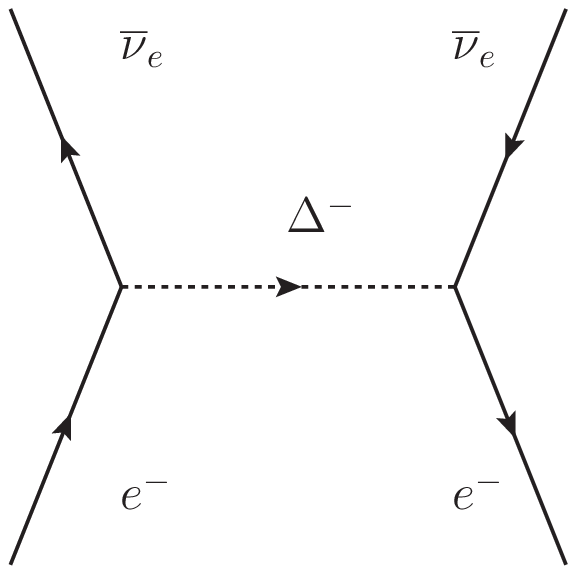}\label{figF3}}
\caption{Feynman diagrams for $\overline{\nu}_e e$ scattering.}
\label{figF}
\end{figure}

The $\overline{\nu}_e-e$ scattering in the frame of HTM proceeds through the exchange of $W^\pm$, $Z^0$ and $H^{\pm }$ bosons. The charged current and neutral current Lagrangian are \cite{1280701}
\begin{equation}
\mathcal{L}_{CC}=-\frac{g}{2\sqrt{2}}\overline e\gamma^\mu(1-
\gamma_5){\nu}_e W^-_\mu+H.c \label{eq4}
\end{equation}
\begin{eqnarray}
\mathcal{L}_{NC}&=&-\frac{g}{4\cos\theta_W}[\overline{\nu}
_e\gamma^\alpha(1-\gamma_5)\nu_e- \notag \\
&& \overline{e}\gamma^\mu(1-4\sin^2\theta_W-\gamma_5)e]Z^0_\mu+H.c
\label{eq5}
\end{eqnarray}
respectively where $g^2=8m_W^2G_F/\sqrt{2}$. The coupling of leptons to $H^{\pm }$ bosons from Eq. (\ref{eq1}), is
\begin{equation}
\mathcal{L} =-h_{ll'}\sqrt{2}\left( l_{l}^{T}CP_{L}\nu_{l'}+\nu_{l}^{T}CP_{L}l_{l'}\right) \cos\theta_+H^{+}+H.c
\end{equation}
The mixing angle $\cos\theta_+$ is approximated to unity due to the suppression of $\upsilon_\Delta/\upsilon_0 \lesssim0.02$ form the $\rho$ parameter [see Eq. (\ref{eq2.49})]. Therefore, the contribution from $H_2$ exchange with the Yukawa coupling $-im_f/\upsilon_0\sin\theta_0$ is zero. Following the convention in \cite{PhysRevD.22.2122}, $t$ is the square of the momentum transferred between incoming $\overline{\nu}_e$ and outgoing singly charged particle (i.e. $e^-$, $\mu^-$, $\Delta^-$). Thus, the Z boson exchange would be denoted as u channel (Fig. \ref{figF}(b)). The scattering angle $\theta$ refers to the angle between incoming $\overline{\nu}_e$ and outgoing singly charged particle in the centre-of-mass frame. The s channels come from $W^-$ (Fig. \ref{figF}(a)) and $%
H^-$ bosons exchange (Fig. \ref{figF}(c)). The
differential scattering cross section is found to be
\begin{widetext}
\begin{eqnarray}
\frac{d\sigma }{dt}&=&\frac{G_{F}^{2}}{\pi s^{2}}\left\{ \frac{m_{W}^{4}}{%
(s-m_{W}^{2})^{2}+m_W^2\Gamma_W^2}t^{2}+\frac{m_{Z}^{4}}{4(s+t+m_{Z}^{2})^{2}}[(2\sin^{2}\theta _{W}-1)^{2}t^{2}+(2\sin^{2}\theta _{W})^{2}s^{2}%
]+ \frac{ m_{W}^{2}m_{Z}^{2}(2\sin^{2}\theta _{W}-1)}{%
(s-m_{W}^{2})(u-m_{Z}^{2})}t^{2}\right. \notag
\\
&& {}\left.+\frac{h_{ee}^{4}}{%
8G_{F}^{2}(s-M_{H^\pm}^{2})^{2}+M_{H^\pm}^{2}\Gamma_{H^\pm}^2}t^{2}\right\} \label{eq6}
\end{eqnarray}
where $t=-(1-\cos\theta)s/2$ and $s=2m_eE$. The differential cross section is integrated over all angle from $\cos\theta=-1$ to $\cos\theta=1$ correspond to $t=-s$ to $t=0$. The total cross section is
\begin{eqnarray}
\sigma_{tot}&=&\frac{G_{F}^{2}s}{\pi} \left[ \frac{m_{W}^{4}}{
3(s-m_{W}^{2})^{2}+m_W^2\Gamma_W^2}+\frac{h_{ee}^{4}}{24G_{F}^{2}(s-M_{H^\pm}^{2})^{2}+M_{H^\pm}^{2}\Gamma_{H^\pm}^2}\right.\nonumber\\& & {}\left.+\frac{1}{4}\left(\frac{m_Z^2}{s}\right)^2[(2%
\sin^{2}\theta _{W}-1)^{2}\left[2+\frac{s}{m_Z^2}-2\left(1+\frac{m_Z^2}{s}
\right)\ln\left(1+\frac{s}{m_Z^2}\right)\right] +(2\sin^{2}\theta _{W})^{2}
\frac{s^2}{m_Z^2(s+m_Z^2)}]\right.\nonumber\\& & {}\left.+\frac{m_W^2m_Z^2}{s(s-m_W^2)}(2\sin^{2}\theta _{W}-1)\left[\frac{3}{2}+\frac{m_Z^2}{s}-\left(1+\frac{m_Z^2}{s}\right)^2\ln\left(1+\frac{m_Z^2}{s}\right)\right]\right] \label{eq7}
\end{eqnarray}
\end{widetext}
where $\Gamma_W$ and $\Gamma_{H^\pm}$ are the total decay width of $W^-$ and $H^-$ boson respectively. The result in Eq. (\ref{eq7}) agree with \cite{PhysRevD.22.2122} except for the second term which comes form $H^-$ exchange. 
\section{The upper bound of $h_{ee}/M_{H^\pm}$}
The upper bound of $h_{ee}/M_{H^\pm}$ can be estimated through low energy $\overline{\nu}_e e$ scattering experiment. The total cross section of Eq. (\ref{eq7}) at low energy can be written in
power series. The terms proportional to $s/m_W, s/M_{H^\pm}, s/m_Z$ are neglected due $s<< M_W^2, M_{H^\pm}^2, M_Z^2$. The
total cross section reduces to
\begin{eqnarray}
\sigma _{tot}&=&\frac{G_{F}^{2}s}{\pi }\left\{ \frac{1}{3}+\frac{(2\sin
^{2}\theta _{W}-1)^{2}}{12}+\frac{(2\sin ^{2}\theta _{W})^{2}}{4}+\right. \notag
\\
&& {}\left.\frac{%
(2\sin ^{2}\theta _{W}-1)}{3}+\frac{h_{ee}^{4}}{24G_{F}^{2}M_{H^\pm}^{4}}%
\right\} \nonumber \\ \label{eq4.1}
\end{eqnarray}
which is the famous neutrino electron elastic scattering at low energy except for the last term which comes from $H^-$ exchange.
The experimental result from the TEXONO collaboration \cite{PhysRevD.81.072001} gives
\begin{equation}
\frac{\sigma_{exp}}{\sigma_{SM}}=1.08\pm0.21\pm0.16 \label{eq4.2}
\end{equation}
By setting $\sigma_{exp}=\sigma_{tot}$, one get 
\begin{equation}
h_{ee}/M_{H^\pm}\lesssim2.8%
\times10^{-3} GeV^{-1} \label{eq10}
\end{equation}
which is close to the limit estimated in \cite{Coarasa1996131} by using the $\nu_e e$ scattering process. The diagonal and non-diagonal upper bound on the coupling constant matrices other than $h_{ee}$ are \cite{PhysRevD.40.1521, JOUR}
\begin{eqnarray}
h_{e\mu}h_{ee}<3.2\times10^{-11}M^2_{\Delta}\nonumber\\
h_{\mu\mu}^2\lesssim2.5\times10^{-5}M^2_{\Delta}\nonumber\\
h_{ee}h_{\mu\mu}\lesssim5.8\times10^{-5}M^2_{\Delta}\nonumber\\
h_{e\mu}h_{\mu\mu}\lesssim2\times10^{-10}M^2_{\Delta}\nonumber\\
\end{eqnarray}
However, the limit on the couplings of third generation leptons at present are poorly known.

\section{Result}
In this section, we present the numerical results of the cross section for $\overline{\nu}_e$ scattering process in the range 0 $\leq E \leq$ 500 GeV, 0.5 TeV $\leq E \leq$ 1000 TeV and $10^{15}$ eV $\leq E \leq$ $10^{20}$ eV. We take these range of energies so that the neutrinos from various possible sources are considered. We set $m_W$ = 80.399 GeV, $m_Z$ =
91.1876 GeV, $G_F=1.166\times10^{-5}$ GeV$^{-2}$, $\Gamma_W=$ 2.085 GeV and $\sin^2\theta_W$ = 0.23116 \cite{09543899377A075021}. The lower bound of the charged Higgs boson mass $M_{\Delta}\gtrsim$110 GeV from the direct search at Tevatron is used  \cite{PhysRevLett.93.141801, *PhysRevLett.93.221802}. The singly charged Higgs boson, $H^-$ decays into final state fermions and final state bosons. The total decay width of $H^-$ is written as \cite{PhysRevD.78.015018}
\begin{eqnarray*}
\Gamma_{H^\pm}&=&\sum_{l, l'}\Gamma (H^-\rightarrow l^-\nu_l)+\sum_{q}\Gamma (H^-\rightarrow \overline{t}b)\nonumber\\&&+\Gamma (H^-\rightarrow W^-_TZ_T)+\Gamma (H^-\rightarrow W^-_LZ_L)\nonumber\\&&+\Gamma (H^-\rightarrow W^-_LH_1)
\end{eqnarray*}
where the subscript T and L stand for the transverse and longitudinal polarization of gauge bosons, respectively. Base on the discussion in Ref. \cite{PhysRevD.78.015018}, leptonic decays of $H^-$ are more favorable for small value of vev. For large value of vev $\upsilon_\Delta\sim$  1 GeV, $H^-$ decays into bosons and quarks which is not very relevant for the process under consideration. Thus, with the choice of $\upsilon_\Delta$ = 1 eV, $H^-$ decays predominantly into leptons and the leptonic decay width is
\begin{equation}
\Gamma (H^-\rightarrow l^-\nu_l)=\frac{\left | h_{ll} \right |^2}{16\pi}M_{H^-}
\end{equation}
The value of 1 eV is the lower bound to the vev based on the naturalness consideration. In this work, we assume $h_{ee}=h_{\mu\mu}=h_{\tau\tau}$ and thus the total decay width is $\Gamma_{H^\pm}=3\times \Gamma (H^-\rightarrow l^-\nu_l)$. We only consider for lepton-number conserving process.

\begin{figure}[ht]
\includegraphics[width=8.5cm]{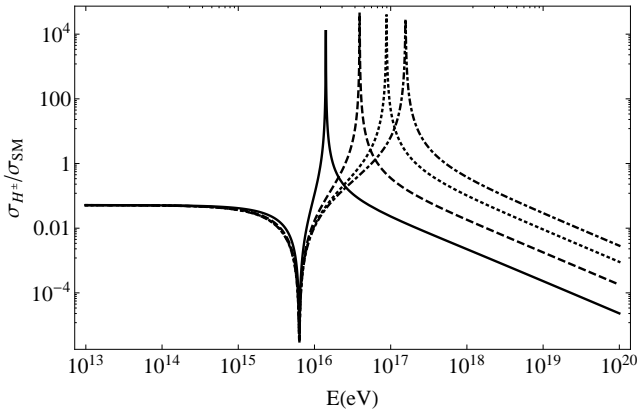}
\caption{The ratio $\sigma_{H^\pm}/\sigma_{SM}$ as the function of $E$ for $h_{ee}=0.30$, $M_{H^\pm}=120$ GeV and $\Gamma_{H^\pm}=0.63$ GeV (solid line); $h_{ee}=0.50$, $M_{H^\pm}=200$ GeV and $\Gamma_{H^\pm}=2.97$ GeV (dashed line); $h_{ee}=0.75$, $M_{H^\pm}=300$ GeV and $\Gamma_{H^\pm}=10.08$ GeV (dotted line); $h_{ee}=1.00$, $M_{H^\pm}=400$ GeV and $\Gamma_{H^\pm}=23.88$ GeV (dash-dotted line).}
\label{fig1b}
\end{figure}

We present the ratio of $\sigma_{H^\pm}/\sigma_{SM}$ in Fig.(\ref{fig1b}) for $h_{ee}=0.30$, $M_{H^\pm}=120$ GeV and $\Gamma_{H^\pm}=0.63$ GeV; $h_{ee}=0.50$, $M_{H^\pm}=200$ GeV and $\Gamma_{H^\pm}=2.97$ GeV; $h_{ee}=0.75$, $M_{H^\pm}=300$ GeV and $\Gamma_{H^\pm}=10.08$ GeV; $h_{ee}=1.00$, $M_{H^\pm}=400$ GeV and $\Gamma_{H^\pm}=23.88$ GeV. Since in all cases $h_{ee}/M_{H^\pm}=2.5%
\times10^{-3}$ GeV$^{-1}$, then the condition (\ref{eq10}) is fulfilled. For the cases considered above, the neutrino mass is $m_{\overline{\nu}_e}<$ 1.5 eV which is below the experimental upper bound $m_{\overline{\nu}_e}<$ 2.3 eV \cite{K2005197} . We take $M_{H^\pm}$ up to 400 GeV so that $h_{ee}\leq1$. At $E\leq10^{14}$ eV, $\sigma_{H^\pm}/\sigma_{SM}\sim0.0516$ for all cases. The $H^-$ boson exchange is almost zero at 6.3 PeV but maximum at their resonance energies.
\begin{figure}[ht]
\centering
\includegraphics[width=8.5cm]{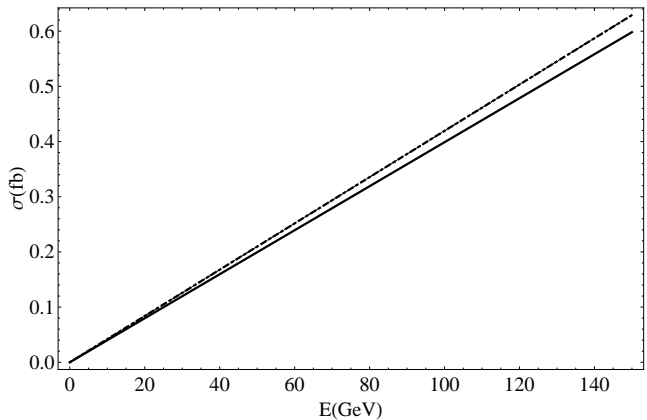}
\caption{Total cross section for $\overline{\nu }_{e}-e$ scattering
in the range 0 $\leq E \leq$ 150 GeV for $h_{ee}=0.30$, $M_{H^\pm}=120$ GeV  and $\Gamma_{H^\pm}=0.63$ GeV; $h_{ee}=0.50$, $M_{H^\pm}=200$ GeV and $\Gamma_{H^\pm}=2.97$ GeV; $h_{ee}=0.75$, $M_{H^\pm}=300$ GeV and $\Gamma_{H^\pm}=10.08$ GeV; $h_{ee}=1.00$, $M_{H^\pm}=400$ GeV and $\Gamma_{H^\pm}=23.88$ GeV. The lines with $H^-$ exchange (dotted) are above the SM prediction (solid line).}
\label{fig:5}
\end{figure}
In Fig. (\ref{fig:5}), we present the total cross section in the range 0 $\leq E \leq$ 150 GeV to show the difference between $H^-$ exchange and SM for all cases. Furthermore, we also present the total cross section as a function of higher energy for $h_{ee}=0.30$, $M_{H^\pm}=120$ GeV and $\Gamma_{H^\pm}=0.63$ GeV; $h_{ee}=0.50$, $M_{H^\pm}=200$ GeV and $\Gamma_{H^\pm}=2.97$ GeV; $h_{ee}=0.75$, $M_{H^\pm}=300$ GeV and $\Gamma_{H^\pm}=10.08$ GeV; $h_{ee}=1.00$, $M_{H^\pm}=400$ GeV and $\Gamma_{H^\pm}=23.88$ GeV in Fig. (\ref{fig:6}).

At low energies limit, the total cross section for all cases grow linearly.  Despite different masses and coupling constants of charged Higgs bosons, there is no significant difference among the total cross sections. However, the charged Higgs exchange do have some contribution with respect to SM. In the range of 0.5 TeV $\leq E \leq$ 1000 TeV, the charged Higgs exchange have a small contribution to the total cross section with respect to SM as shown in Fig. \ref{fig:6a}. The first peak in Fig. (\ref{fig:6b}) correspond to the Glashow resonance \cite{PhysRev.118.316} and following peaks are $H^-$ boson resonance for different cases. Clear signals of the resonance peaks can be obtained as the total cross section for the processes are greater than the background for neutrino with energies between $10^{16}$ eV $\leq E \leq10^{19}$ eV.  
\begin{figure}[ht]
\centering
\subfloat[]{\label{fig:6a}\includegraphics[width=8.5cm]{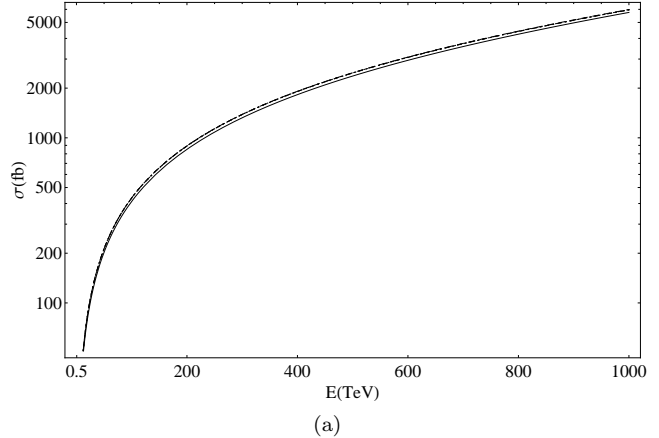}}\\
\subfloat[]{\label{fig:6b}\includegraphics[width=8.6cm]{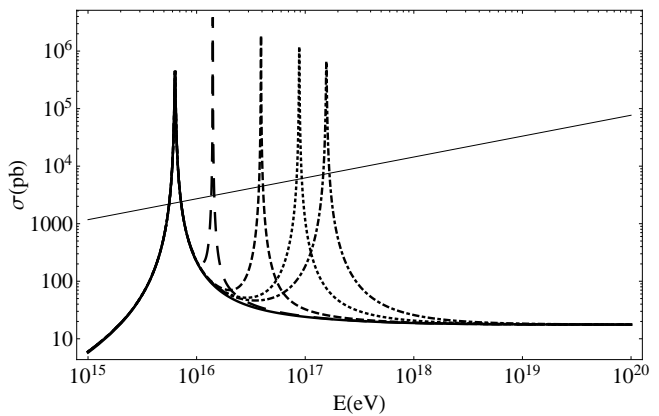}}
\caption{Total cross section for $\overline{\nu}_{e}-e$ scattering
in the range (a)  0.5 TeV $\leq E \leq$ 1000 TeV for all cases. The solid line corresponds to SM. The dashed lines overlap above the SM are $H^-$ exchange for different cases. (b) $10^{15}$ eV $\leq E \leq$ $10^{20}$ eV with SM (solid line), $h_{ee}=0.30$, $M_{H^\pm}=120$ GeV and $\Gamma_{H^\pm}=0.63$ GeV (dashed line); $h_{ee}=0.50$, $M_{H^\pm}=200$ GeV and $\Gamma_{H^\pm}=2.97$ GeV (small-dashed line); $h_{ee}=0.75$, $M_{H^\pm}=300$ GeV and $\Gamma_{H^\pm}=10.08$ GeV (dotted line); $h_{ee}=1.00$, $M_{H^\pm}=400$ GeV and $\Gamma_{H^\pm}=23.88$ GeV (dash-dotted line). The straight line is the cross section for $\overline{\nu}_lN$ deep inelastic scattering given in Ref. \cite{PhysRevD.58.093009} for energies between $10^{16}$ eV $\leq E \leq$ $10^{21}$ eV.}
\label{fig:6}
\end{figure}

\section{Conclusion}
We have presented the $\overline{\nu}_e-e$ scattering cross section in the range 0 $\leq E \leq$ $10^{20}$ eV for 0.30 $\leq h_{ee} \leq$ 1 and 120 GeV $\leq M_{H^\pm} \leq$ 400 GeV. From the low energy $\overline{\nu}_e-e$ scattering experiment, the upper bound $h_{ee}/M_{H^\pm}\lesssim2.8%
\times10^{-3}$ GeV$^{-1}$ was derived. The $H^-$  exchange can contribute up to 5.16 $\%$ to the total cross section at $E\leq10^{14}$ eV. The contribution is maximum at $s\approx M_{H^-}^2$. We observed that the discrimination between the SM and the HTM predictions for the process under consideration can be significant in the interval $10^{16}$ eV $\leq E \leq10^{19}$ eV. 
\begin{acknowledgments}
This work was supported in part
by the Institute of Research Management and Monitoring University of Malaya,
Grant No. PS310/2009C.
\end{acknowledgments}
\bibliography{reference} 
\end{document}